\documentclass{svmult}

\usepackage{makeidx}     % allows index generation
\usepackage{graphicx}    % standard LaTeX graphics tool
\usepackage{multicol}    % used for the two-column index
\makeindex             % used for the subject index
\begin{document}

\title*{Deterministic and stochastic influences on Japan and US stock and foreign 
exchange markets. A Fokker-Planck approach} \toctitle{Deterministic and 
stochastic influences on Japan and US stock and foreign exchange markets. A 
Fokker-Planck approach}

\titlerunning{Deterministic and stochastic influences on ... markets}

\author{Kristinka Ivanova\inst{1} \and Marcel Ausloos\inst{2}  \and Hideki 
Takayasu\inst{3}}

\authorrunning{ K. Ivanova,  M. Ausloos and H. Takayasu}

\institute{ Pennsylvania State University, University Park PA 16802, USA \and 
GRASP, B5, University of Li$\grave e$ge, B-4000 Li$\grave e$ge, Euroland\and Sony 
Computer Science Laboratories, Tokyo 141-0022, Japan}

\maketitle

\begin{abstract} The evolution of the probability distributions of Japan and US 
major market indices, NIKKEI~225 and NASDAQ composite index, and $JPY$/$DEM$ and 
$DEM$/$USD$ currency exchange rates is described by means of the Fokker-Planck 
equation (FPE). In order to distinguish and quantify the deterministic and random 
influences on these financial time series we perform a statistical analysis of 
their increments $\Delta x(\Delta(t))$ distribution functions for different time 
lags $\Delta(t)$. From the probability distribution functions at various 
$\Delta(t)$, the Fokker-Planck equation for $p(\Delta x(t), \Delta(t))$ is 
explicitly derived. It is written in terms of a drift and a diffusion 
coefficient. The Kramers-Moyal coefficients, are estimated and found to have a 
simple analytical form, thus leading to a simple physical interpretation for both 
drift $D^{(1)}$ and diffusion $D^{(2)}$ coefficients. The Markov nature of the 
indices and exchange rates is shown and an apparent difference in the NASDAQ 
$D^{(2)}$ is pointed out.

\end{abstract}


\section{Introduction}

Recent studies have shown that the power spectrum of the stock market 
fluctuations is inversely proportional to the frequency on some power, which 
points to self-similarity in time for processes underlying the market 
\cite{peters1,mantegnastanleybook}. Our knowledge of the random and/or 
deterministic character of those processes is however limited. One rigorous way 
to sort out the noise from the deterministic components is to examine in details 
correlations at $different$ scales through the so called master equation, i.e. 
the Fokker-Planck equation (and the subsequent Langevin equation) for the 
probability distribution function ($pdf$) of signal increments \cite{friedrich}. 
This theoretical approach, so called solving the inverse problem, based on 
$rigorous$ statistical principles \cite{Ernst,ref15}, is often the first step in 
sorting out the $best$ model(s). In this paper we derive FPE, directly from the 
experimental data of two financial indices and two exchange rates series, in 
terms of a drift $D^{(1)}$ and a diffusion $D^{(2)}$ coefficient. We would like 
to emphasize that the method is model independent. The technique allows 
examination of long and short time scales {\it on the same footing}. The so found 
analytical form of both drift $D^{(1)}$ and diffusion $D^{(2)}$ coefficients has 
a simple physical interpretation, reflecting the influence of the deterministic 
and random forces on the examined market dynamics processes. Placed into a 
Langevin equation, they could allow for some first step forecasting.

\section{Data}

We consider the daily closing price $x(t)$ of two major financial indices, 
NIKKEI~225 for Japan and NASDAQ composite for US, and daily exchange rates 
involving currencies of Japan, US and Europe, $JPY$/$DEM$ and $DEM$/$USD$ from 
January 1, 1985 to May 31, 2002. Data series of NIKKEI~225 (4282 data points) and 
NASDAQ composite (4395 data points) and are downloaded from the Yahoo web site 
($http://finance.yahoo.com/$). The exchange rates of $JPY$/$DEM$ and $DEM$/$USD$ 
are downloaded from $http://pacific.commerce.ubc.ca/xr/$ and both consists of 
4401 data points each. Data are plotted in Fig. 1(a-d). The $DEM$/$USD$ case was 
studied in \cite{friedrich} for the 1992-1993 years. See also [6], [8-10] and 
[11] for some related work and results on such time series signals, some on high 
frequency data, and for different time spans.

\begin{figure} \centering \includegraphics[width=.48\textwidth]{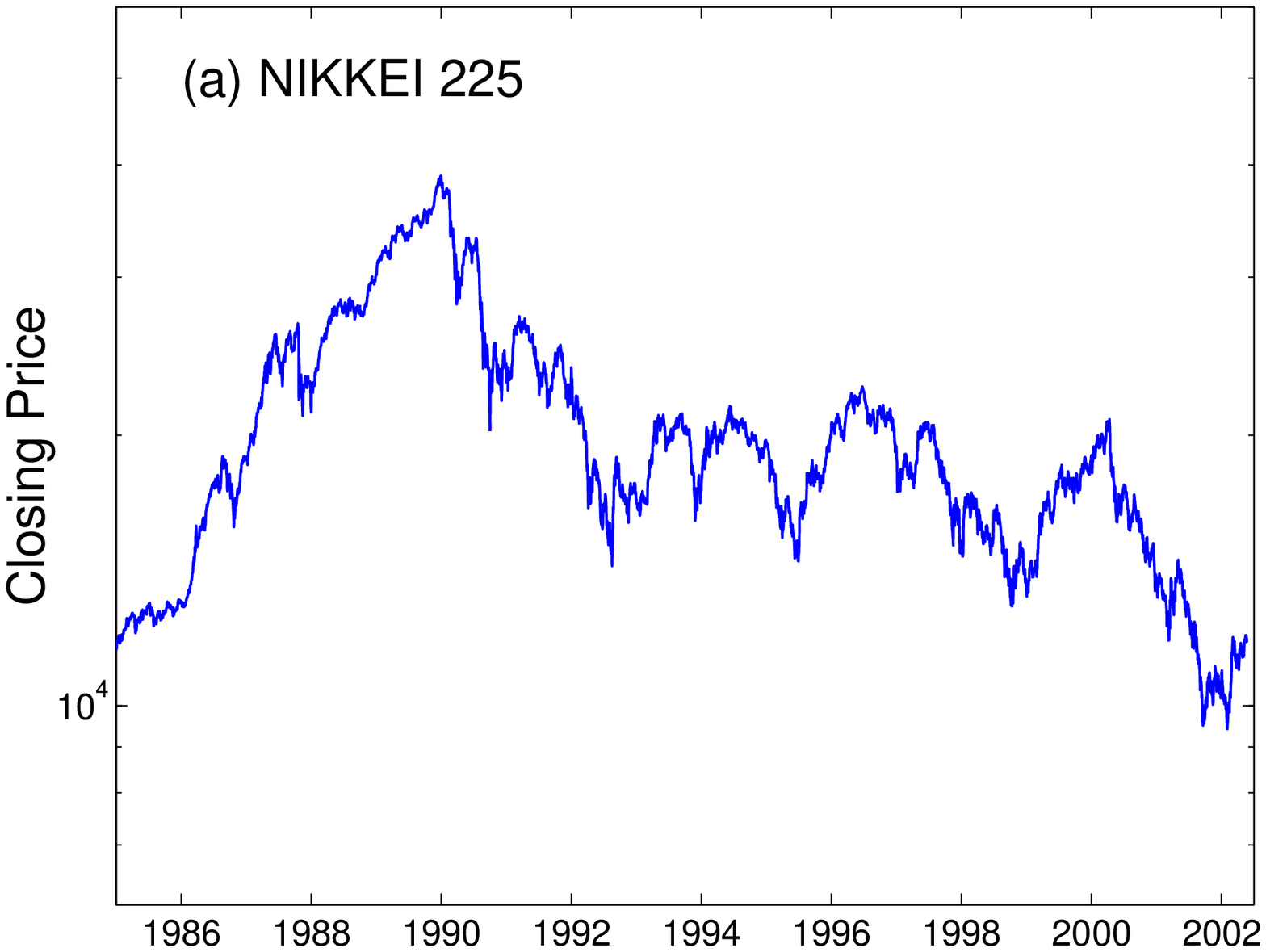} 
\hfill \includegraphics[width=.46\textwidth]{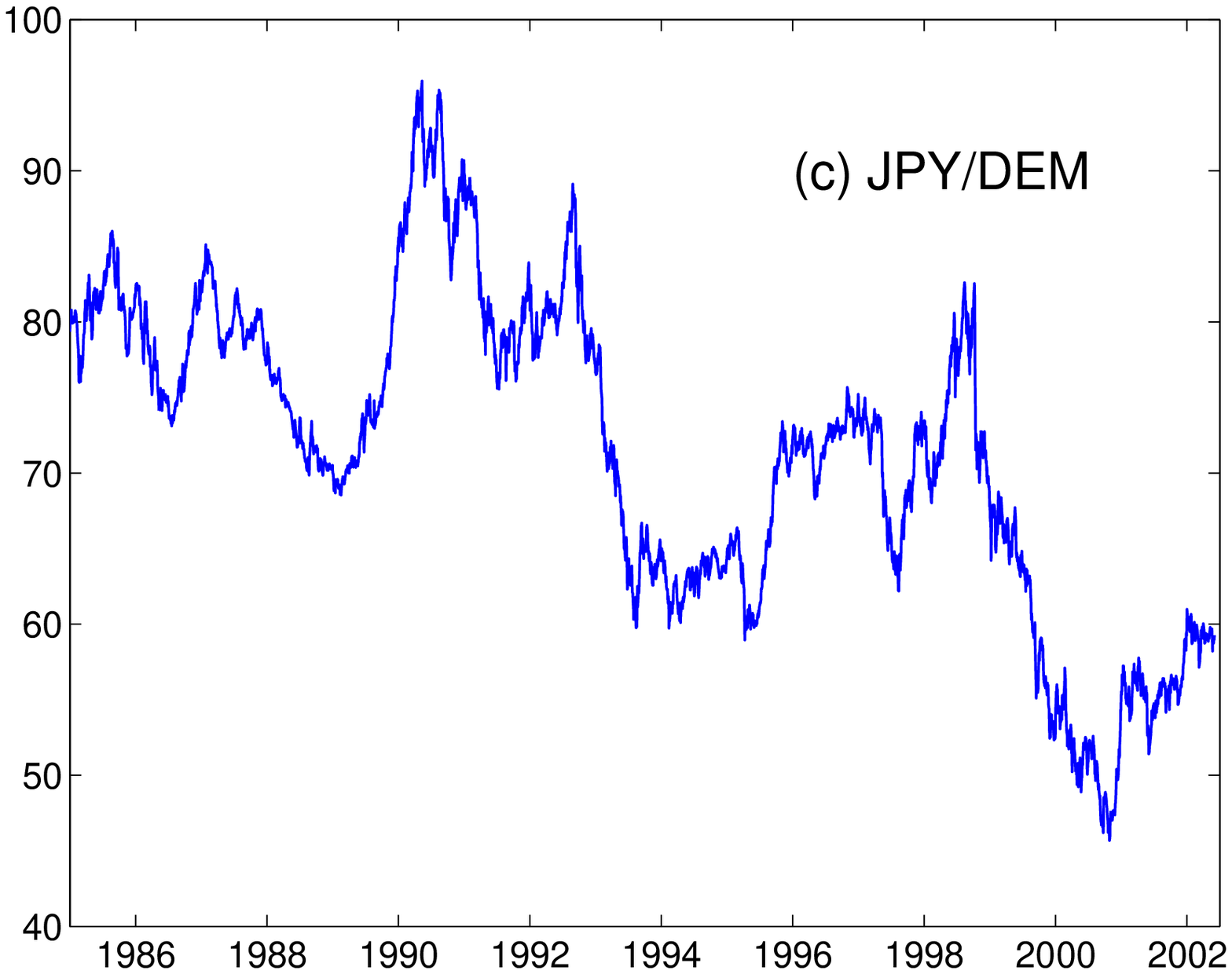} \vfill 
\includegraphics[width=.48\textwidth]{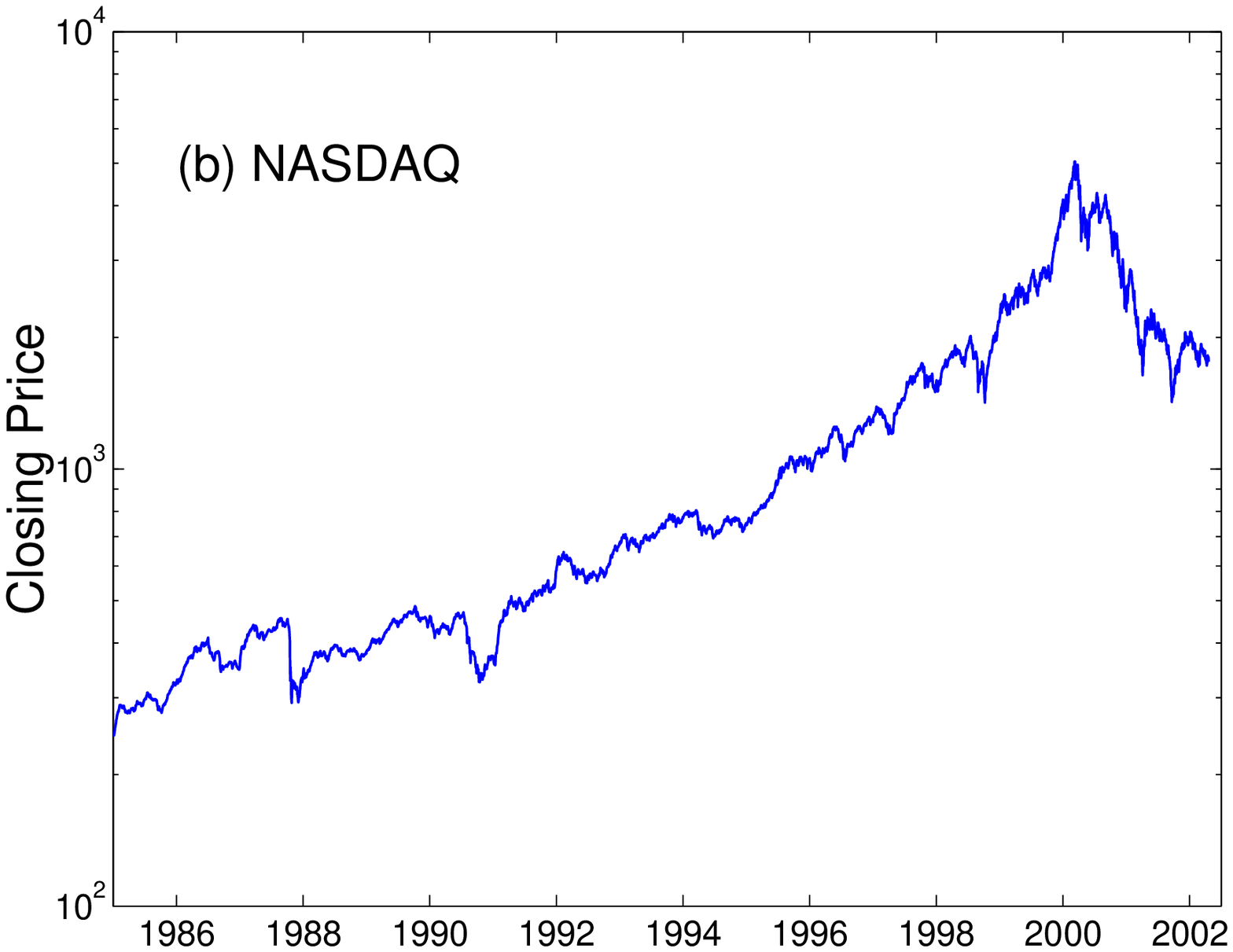} \hfill 
\includegraphics[width=.46\textwidth]{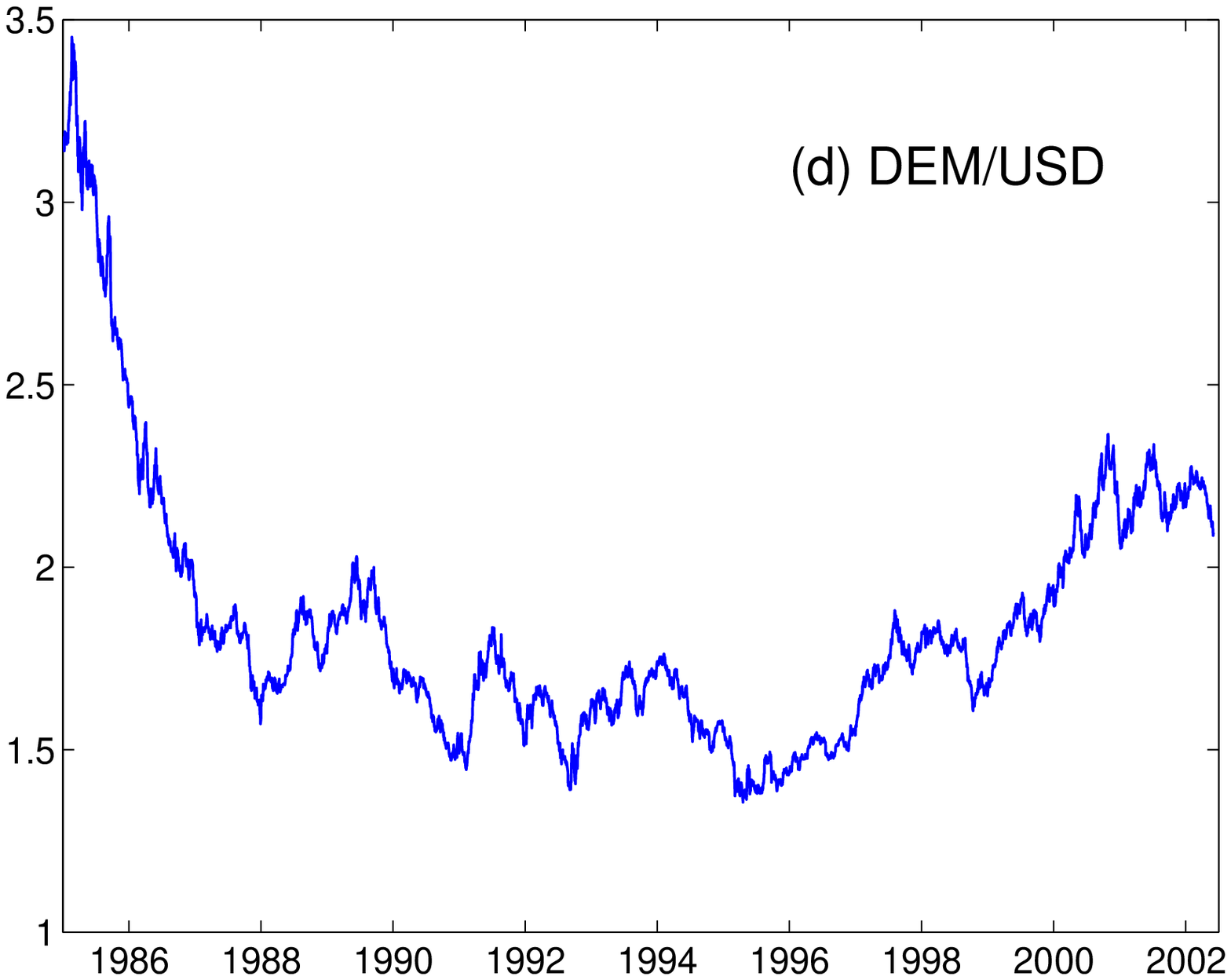} \caption{Daily closing 
price of (a) NIKKEI~225, (b) NASDAQ, (c) $JPY$/$DEM$ and (d) $DEM$/$USD$ exchange 
rates for the period from Jan. 01, 1985 till May 31, 2002} \label{eps1} 
\end{figure}

\section{Results and discussion}

To examine the fluctuations of the time series at different time delays (or time 
lags) $\Delta t$ we study the distribution of the increments $\Delta x=x(t+\Delta 
t)-x(t)$. Therefore, we can analyze the fluctuations at long and short time 
scales on the same footing. Results for the probability  distribution functions 
(pdf) $p(\Delta x,\Delta t)$ are plotted in Fig. 2(a-d). Note that while the pdf 
of one day time delays (circles) for all time series studied have similar shapes, 
the pdf  for longer time delays shows fat tails as in \cite{mantegnastanleybook} 
of the same type for NIKKEI~225, $JPY$/$DEM$ and $DEM$/$USD$, but is different 
from the pdf for NASDAQ.

\begin{figure} \centering \includegraphics[width=.48\textwidth]{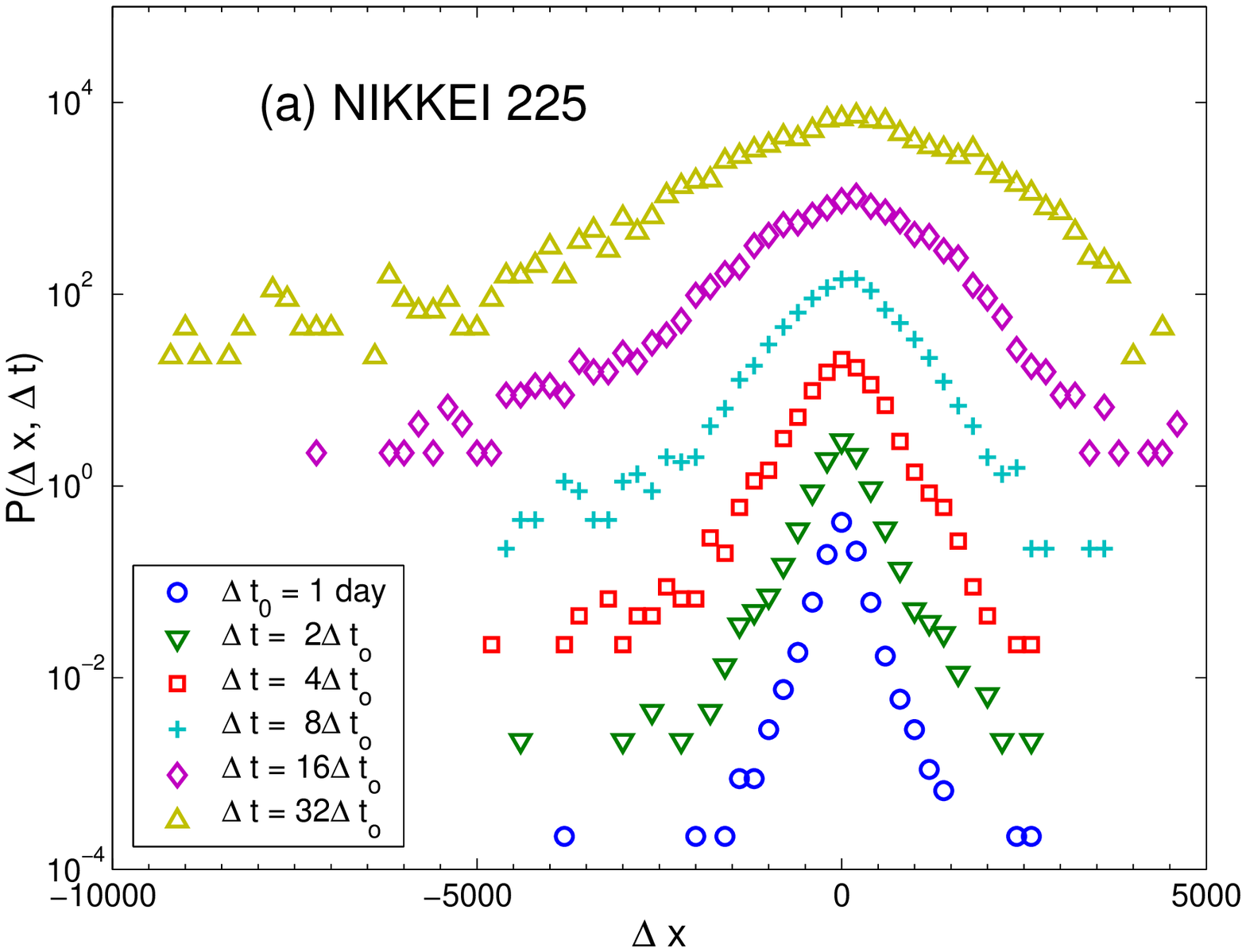} 
\hfill \includegraphics[width=.48\textwidth]{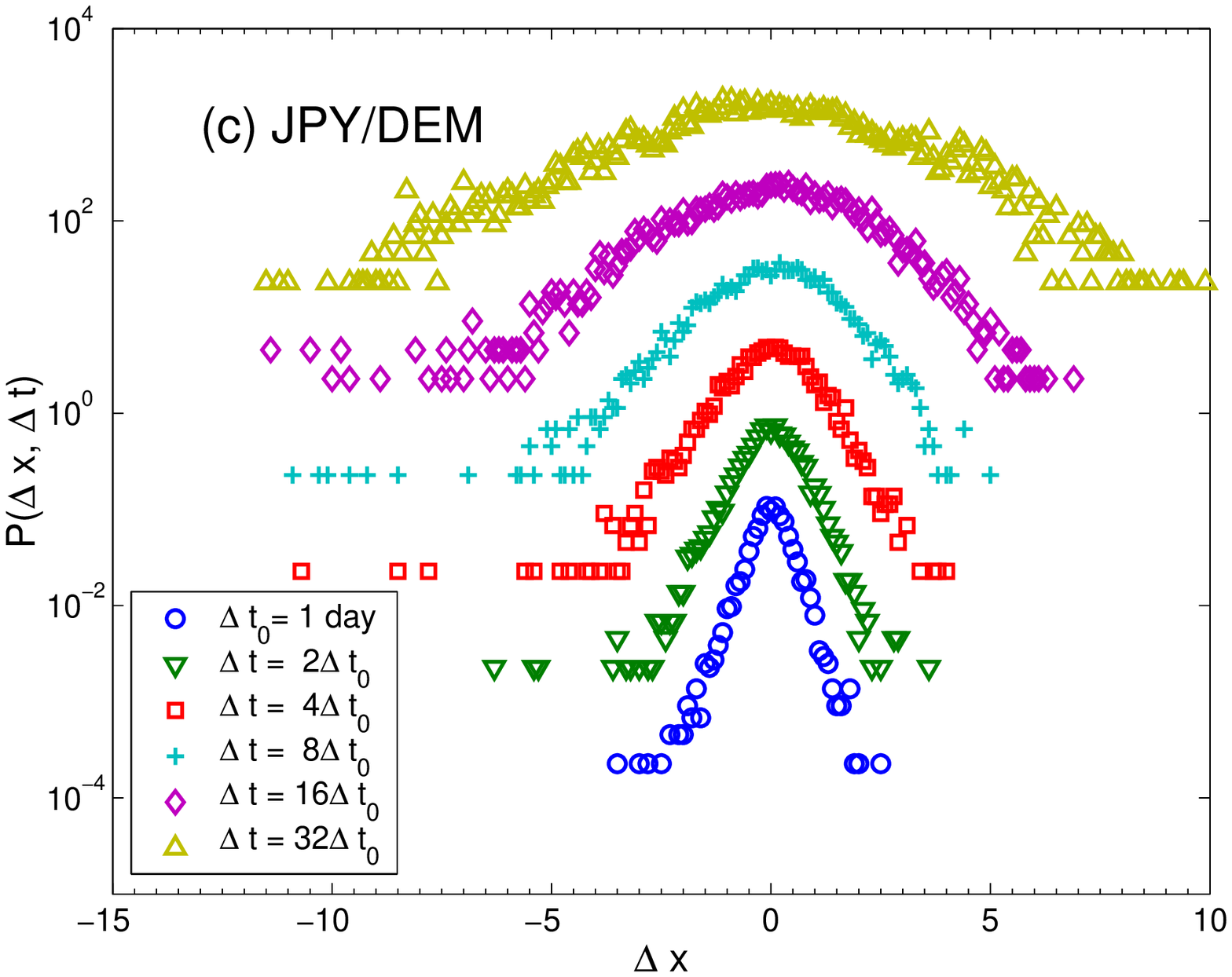} \vfill 
\includegraphics[width=.48\textwidth]{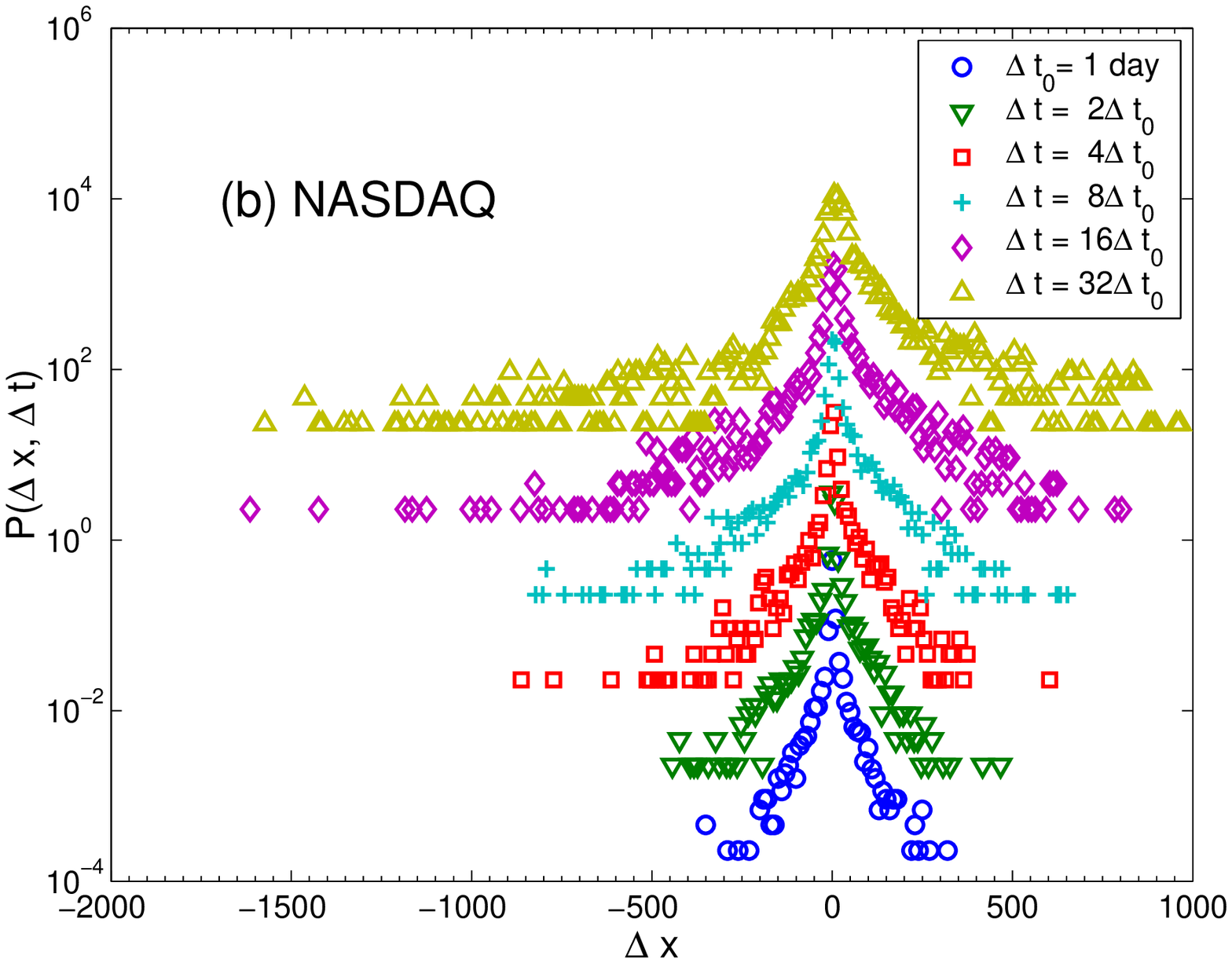} \hfill 
\includegraphics[width=.48\textwidth]{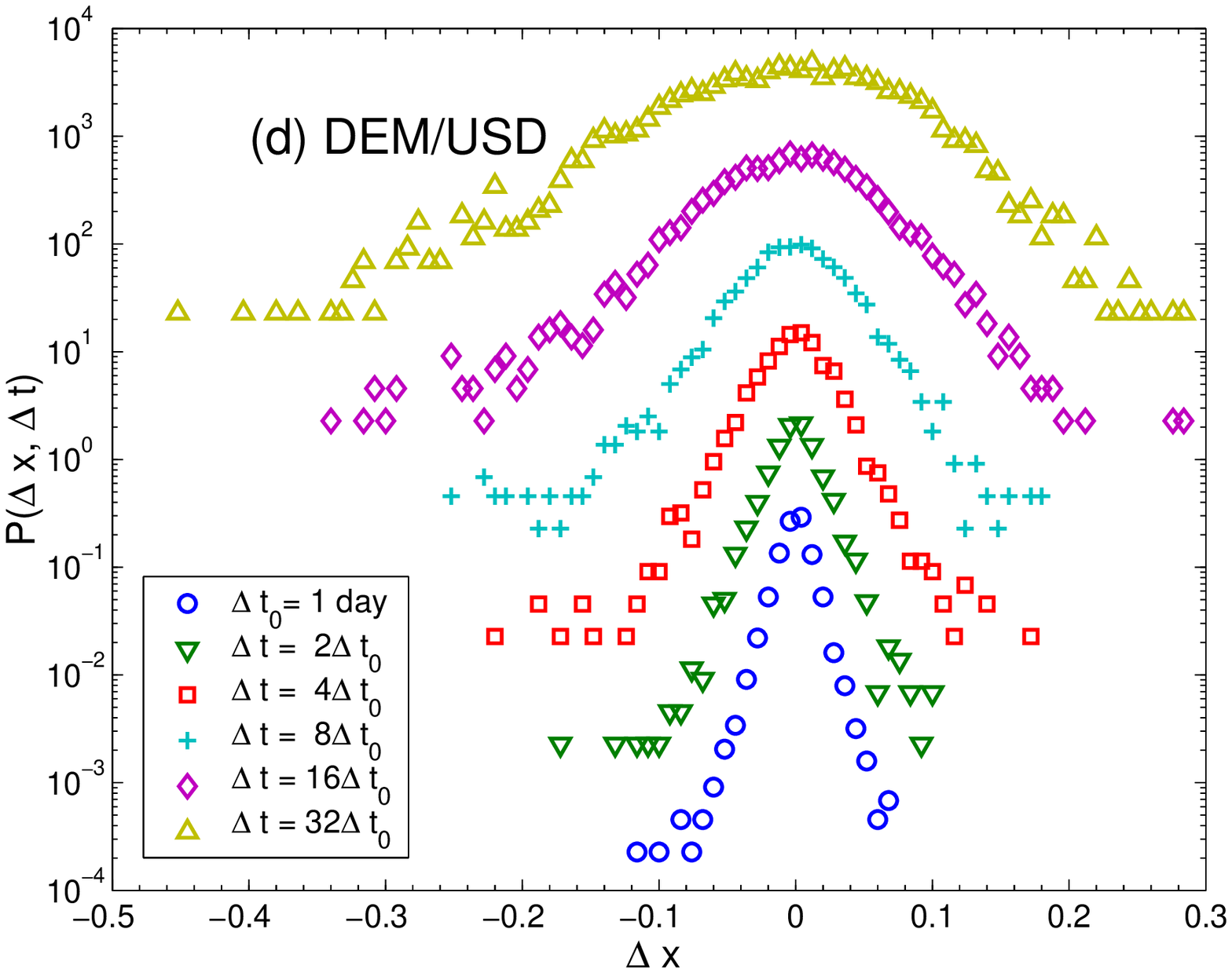} \caption{Probability 
distribution function $p(\Delta x,\Delta t)$ of (a) NIKKEI~225, (b) NASDAQ, (c) 
$JPY$/$DEM$ and (d) $DEM$/$USD$ from Jan. 01, 1985 till May 31, 2002 for 
different delay times. Each pdf is displaced vertically to enhance the tail 
behavior; symbols and the time lags $\Delta t$ are in the insets. The 
discretisation step of the histogram is (a) 200, (b) 27, (c) 0.1 and (d) 0.008 
respectively} \label{eps2} \end{figure}

More information about the correlations present in the time series is given by 
joint pdf's, that depend on $N$ variables, i.e. $p^N (\Delta x_1,\Delta 
t_1;...;\Delta x_N,\Delta t_N)$. We started to address this issue by determining 
the properties of the joint pdf for $N=2$, i.e. $p(\Delta x_2,\Delta t_2; \Delta 
x_1, \Delta t_1)$. The symmetrically tilted character of the joint pdf contour 
levels (Fig. 3(a-c)) around an inertia axis with slope 1/2  points out to the 
statistical dependence, i.e. a correlation, between the increments in all 
examined time series.

\begin{figure} \centering \includegraphics[width=.48\textwidth]{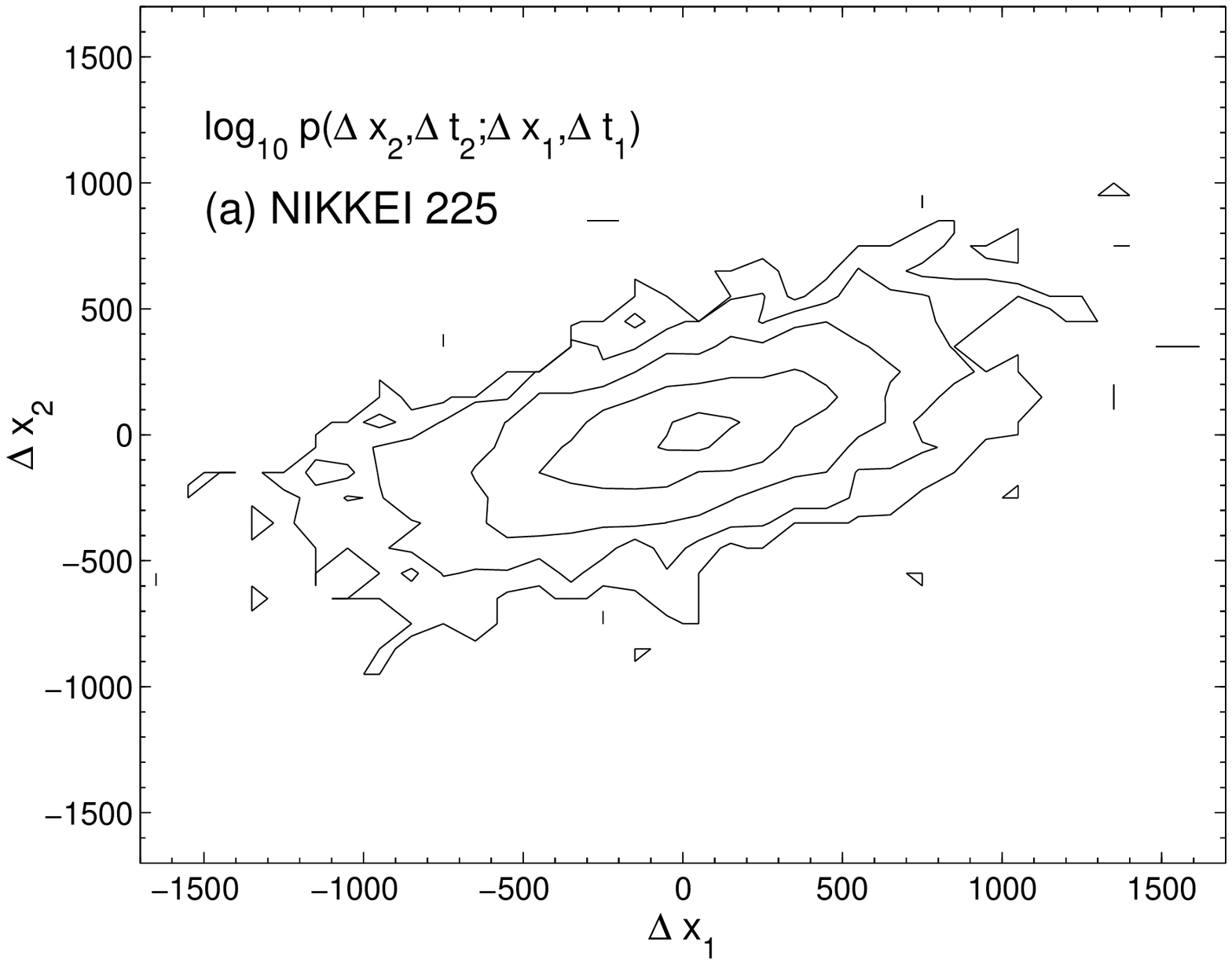} 
\hfill \includegraphics[width=.46\textwidth]{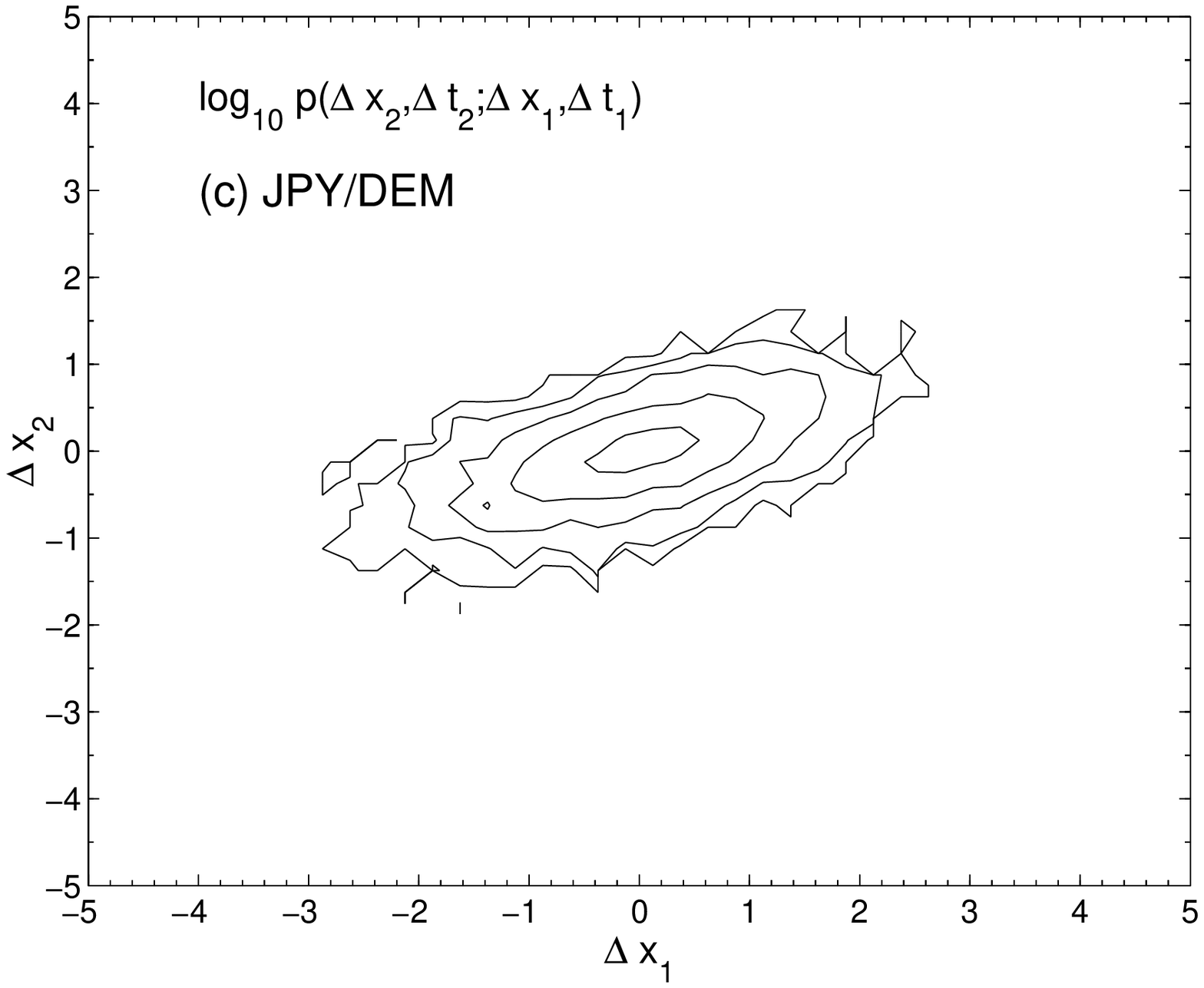} \vfill 
\includegraphics[width=.47\textwidth]{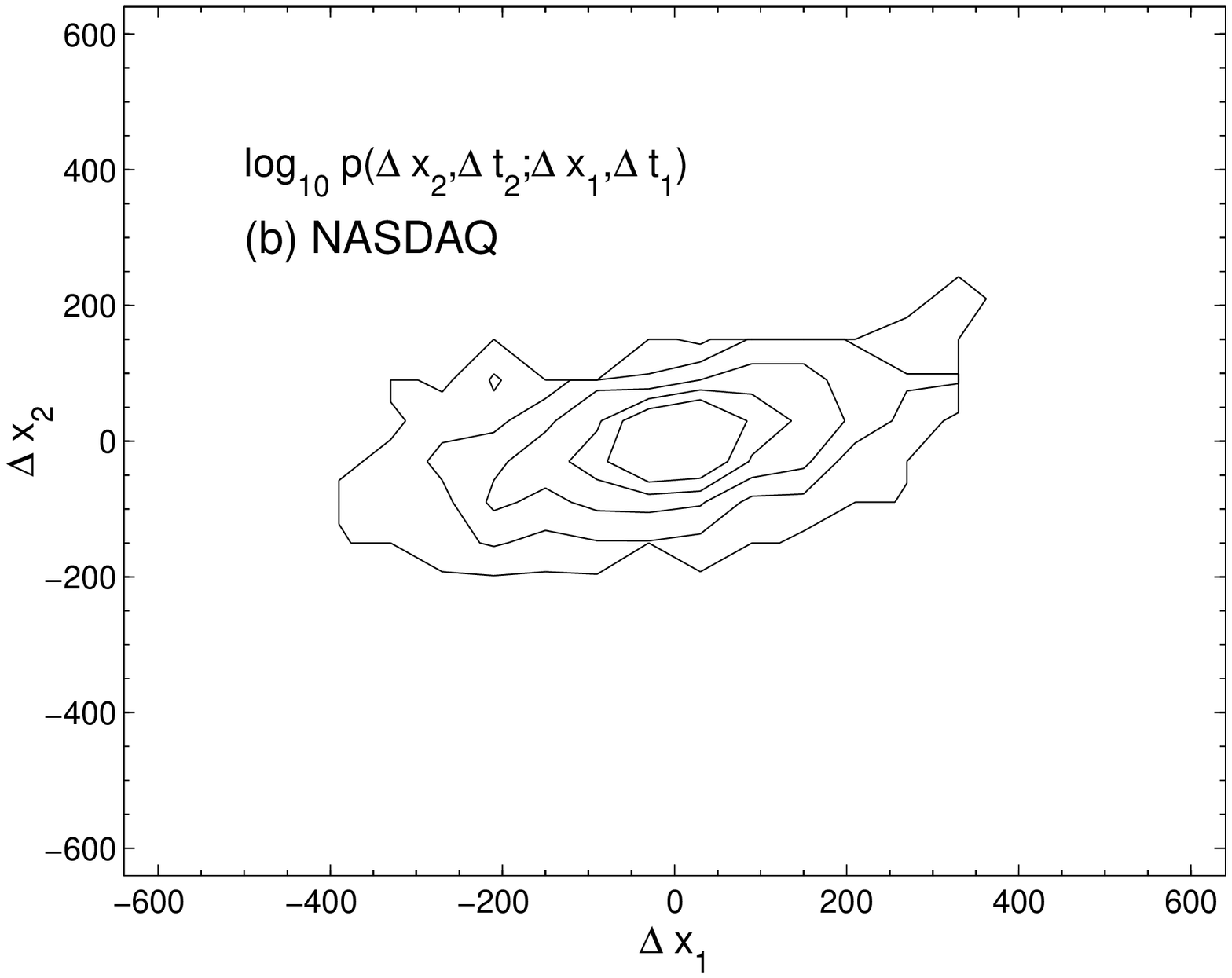} \hfill 
\includegraphics[width=.48\textwidth]{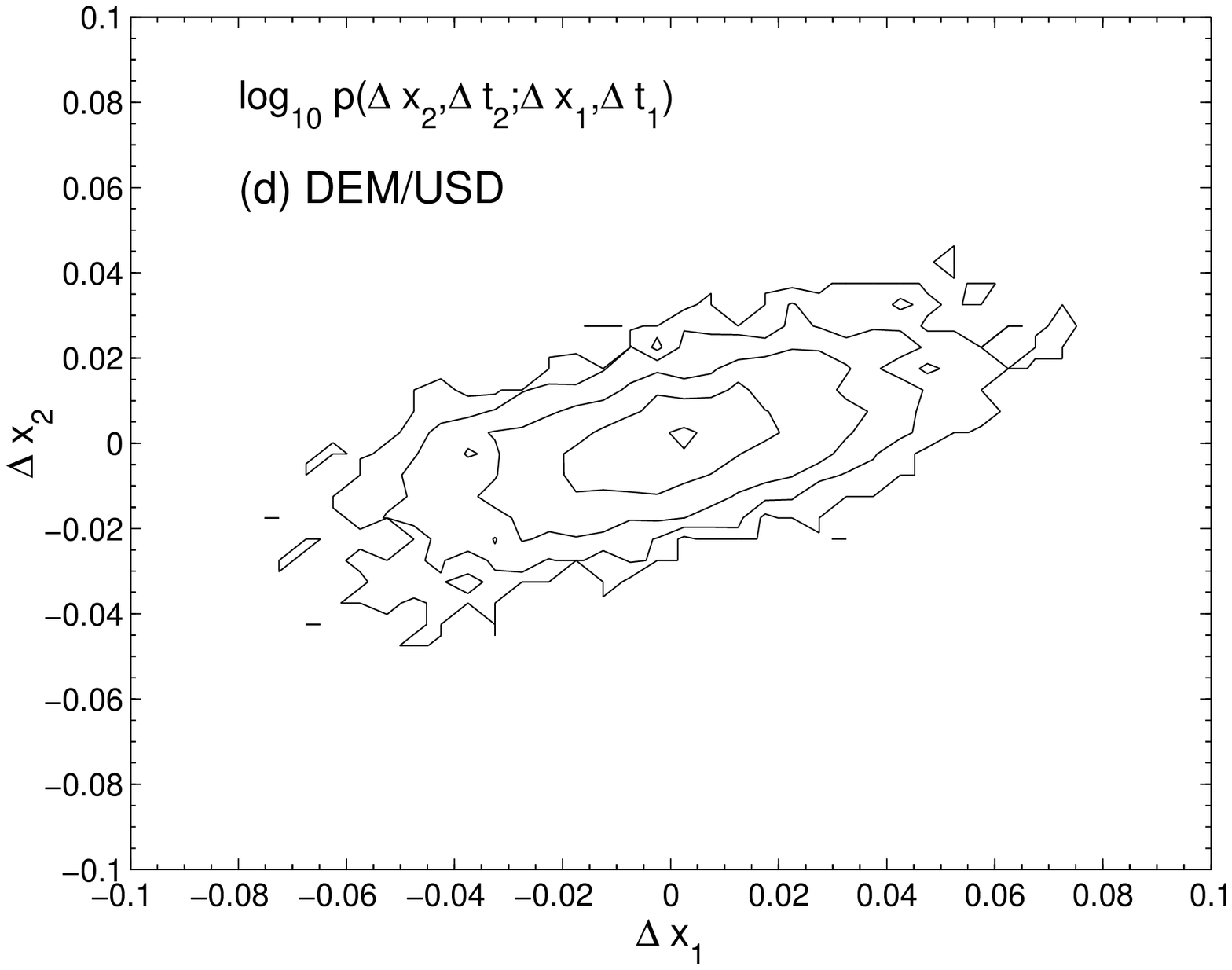} \caption{Typical 
contour plots of the joint probability  density function $p(\Delta x_2,\Delta 
t_2; \Delta x_1,\Delta t_1)$ of (a) NIKKEI~225, (b) NASDAQ closing price signal 
and (c) $JPY$/$DEM$ and (d) $DEM$/$USD$ exchange rates for $\Delta t_2=1\quad 
day$ and $\Delta t_1=3\quad days$. Contour levels correspond to 
$log_{10}p=-1.5,-2.0,-2.5,-3.0,-3.5$ from center to border} \label{eps3} 
\end{figure}

The conditional probability function is

\begin{equation} p ( \Delta x_{i+l},\Delta t_{i+l}|\Delta x_{i},\Delta t_{i}) = 
\frac{p(\Delta x_{i+l},\Delta t_{i+l};\Delta x_{i},\Delta t_{i})}{p(\Delta x_{i}, 
\Delta t_{i})} \end{equation} for $i = 1,...,N-1$.   For any $\Delta t_{2}$ $<$ 
$\Delta t_{i}$ $<$ $\Delta t_{1}$, the Chapman-Kolmogorov equation is a necessary 
condition of a Markov process, one without memory but governed by probabilistic 
conditions

\begin{equation} p(\Delta x_{2},\Delta t_{2}|\Delta x_{1},\Delta t_{1})= \int 
d(\Delta x_{i})p(\Delta x_{2},\Delta t_{2}|\Delta x_{i},\Delta t_{i})p(\Delta 
x_{i},\Delta t_{i}|\Delta x_{1},\Delta t_{1}). \end{equation}

The Chapman-Kolmogorov equation when formulated in $differential$ form yields a 
master equation, which can take the form of a Fokker-P1anck equation 
\cite{Ernst}. For $\tau=log_2(32/\Delta t)$,

\begin{equation} \frac{d}{d\tau}p(\Delta x,\tau )=\left[-\frac{\partial 
}{\partial \Delta x}  D^{(1)}(\Delta x,\tau )+\frac{\partial }{\partial 
^{2}\Delta x^{2}} D^{(2)}(\Delta x,\tau )\right]p(\Delta x,\tau ) \label{efp} 
\end{equation} in terms of a drift $D^{(1)}$($\Delta x$,$\tau $) and a diffusion 
coefficient $D^{(2)}$($\Delta x$,$\tau $) (thus values of $\tau $ represent $ 
\Delta t_{i}$, $i=1,...$).

The coefficient functional dependence can be estimated directly from the moments 
$M^{(k)}$ (known as Kramers-Moyal coefficients) of the conditional probability 
distributions:

\begin{equation} M^{(k)}=\frac{1}{\Delta \tau }\int d\Delta x^{^{\prime }}(\Delta 
x^{^{\prime }}-\Delta x)^{k}p(\Delta x^{^{\prime }},\tau +\Delta \tau |\Delta 
x,\tau ) \end{equation}

\begin{equation} D^{(k)}(\Delta x,\tau )=\frac{1}{k!}\mbox{lim} M^{(k)} 
\end{equation} for $\Delta \tau \rightarrow 0$. The functional dependence of the 
drift and diffusion coefficients $D^{(1)}$ and $D^{(2)}$ for the normalized 
increments $\Delta x$ is well represented by a line and a parabola, respectively. 
The values of the polynomial coefficients are summarized in Table 1 and Fig. 4.

\begin{figure} \centering \includegraphics[width=.445\textwidth]{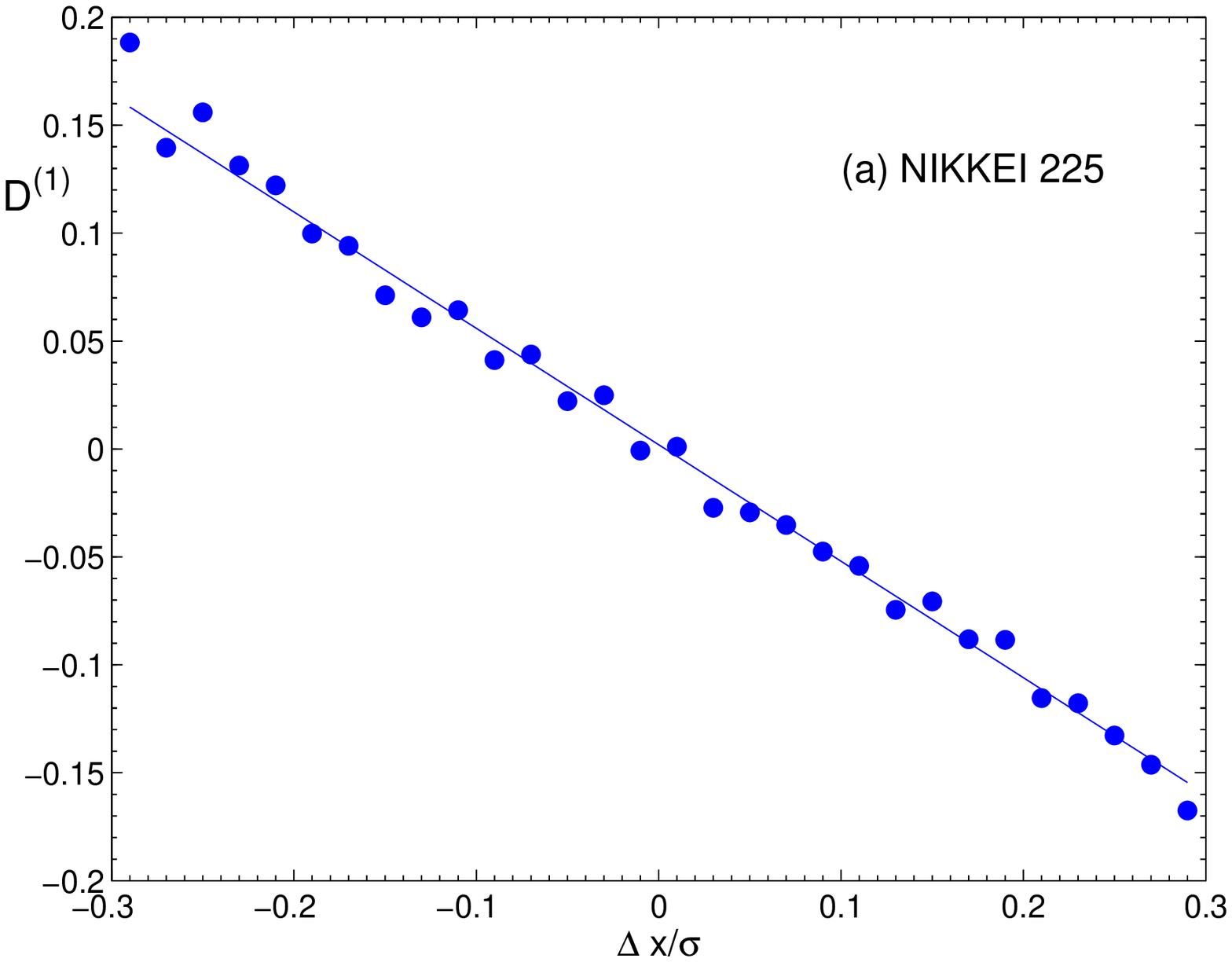} 
\hfill \includegraphics[width=.445\textwidth]{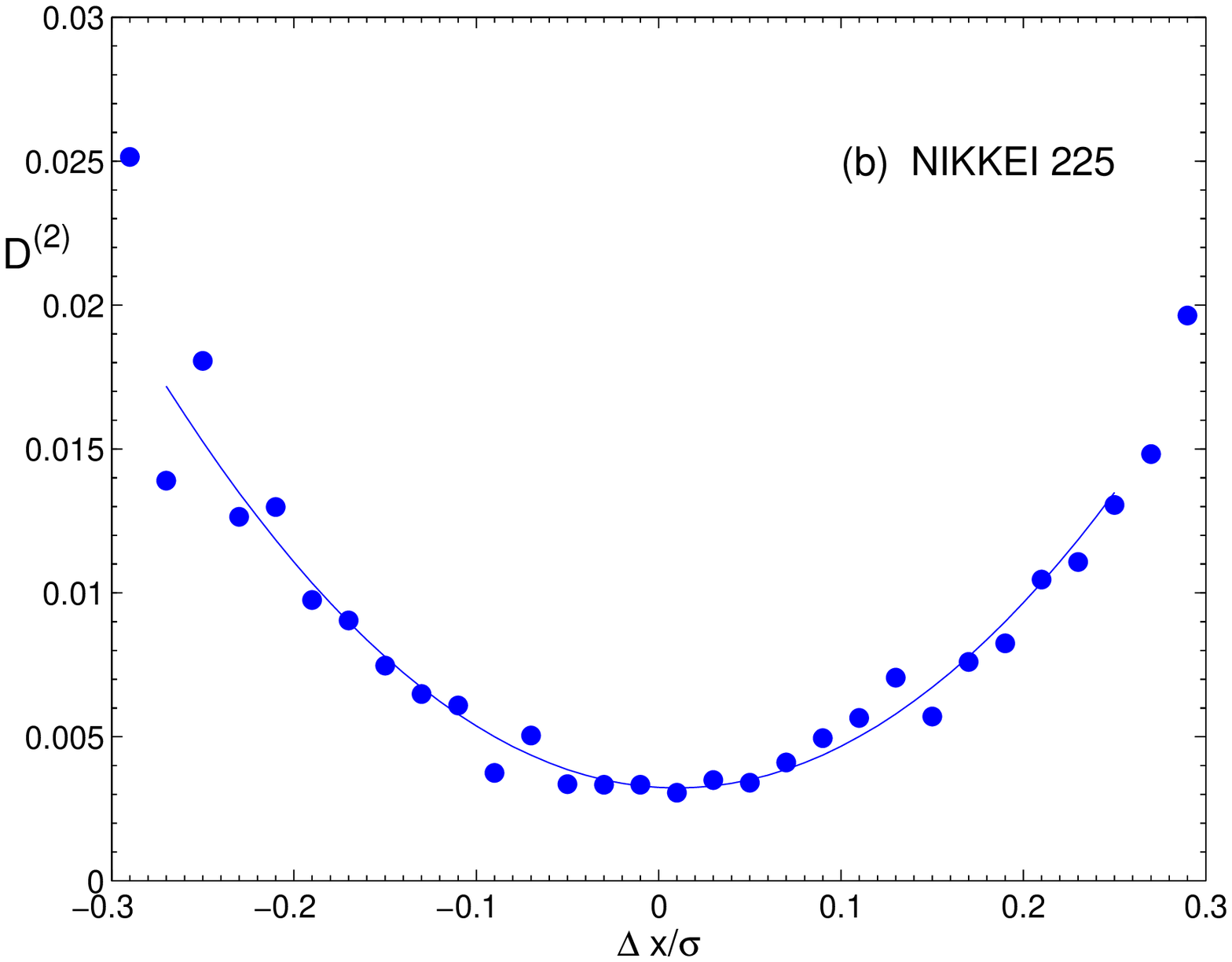} \vfill 
\includegraphics[width=.445\textwidth]{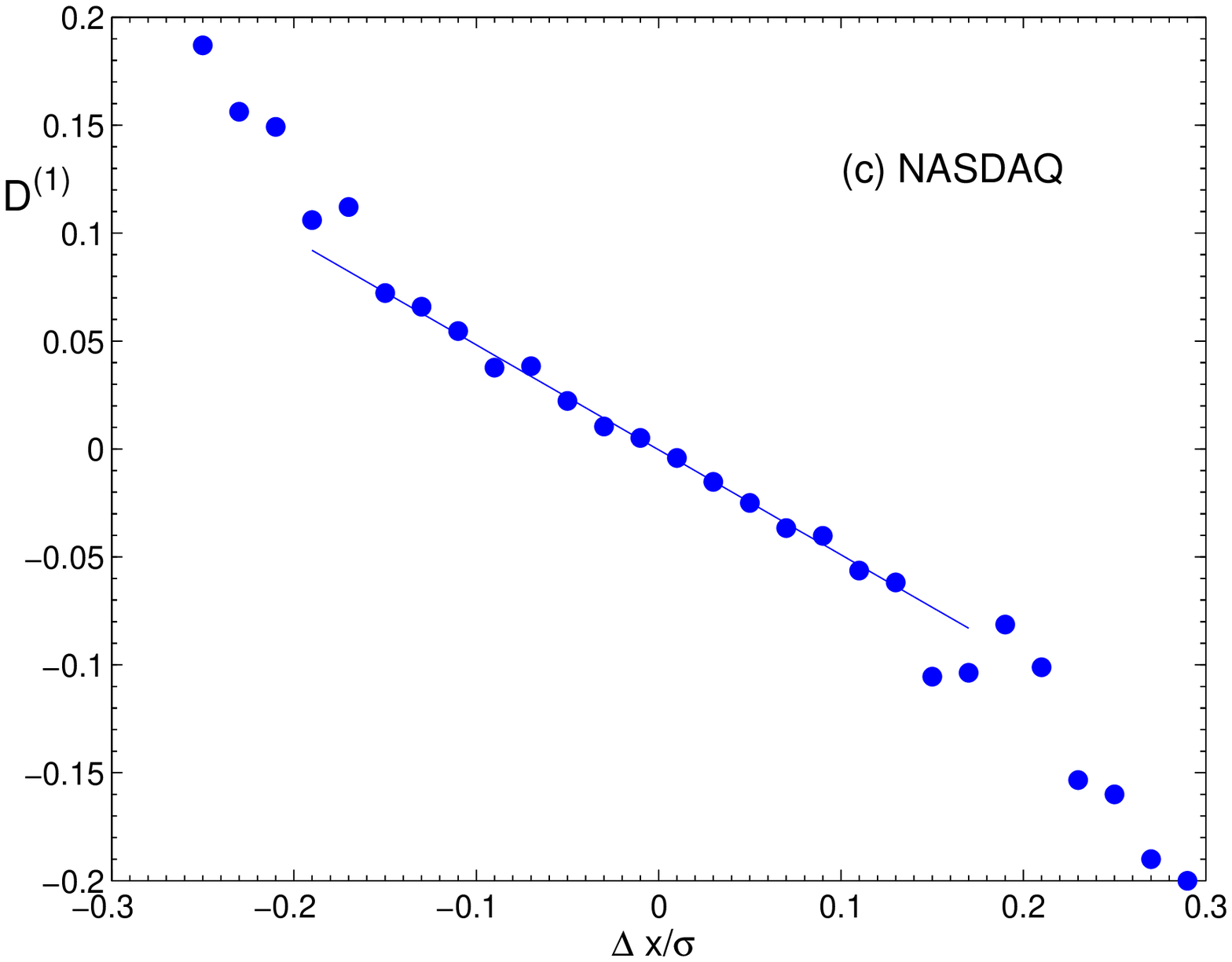} \hfill 
\includegraphics[width=.445\textwidth]{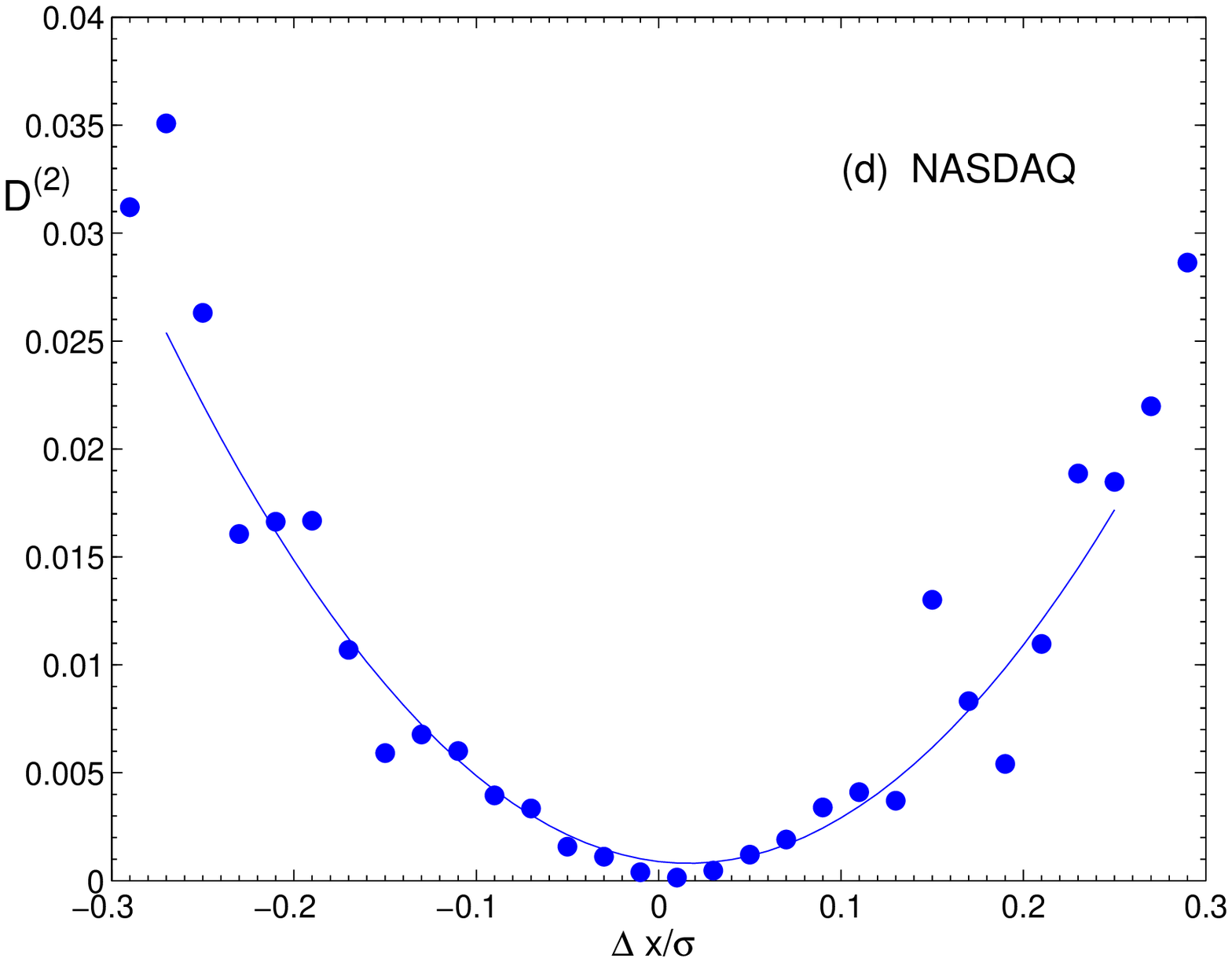} \vfill 
\includegraphics[width=.445\textwidth]{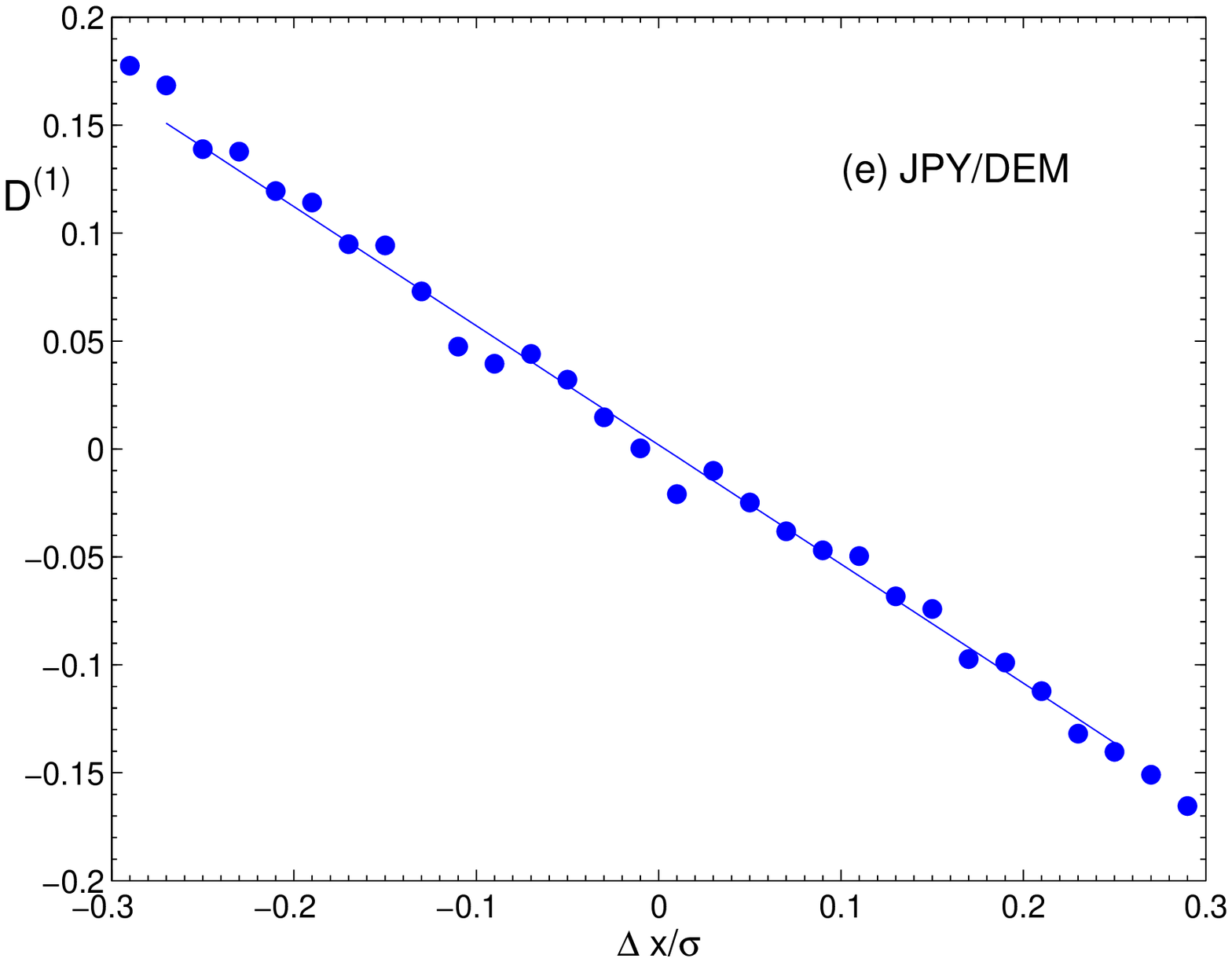} \hfill 
\includegraphics[width=.445\textwidth]{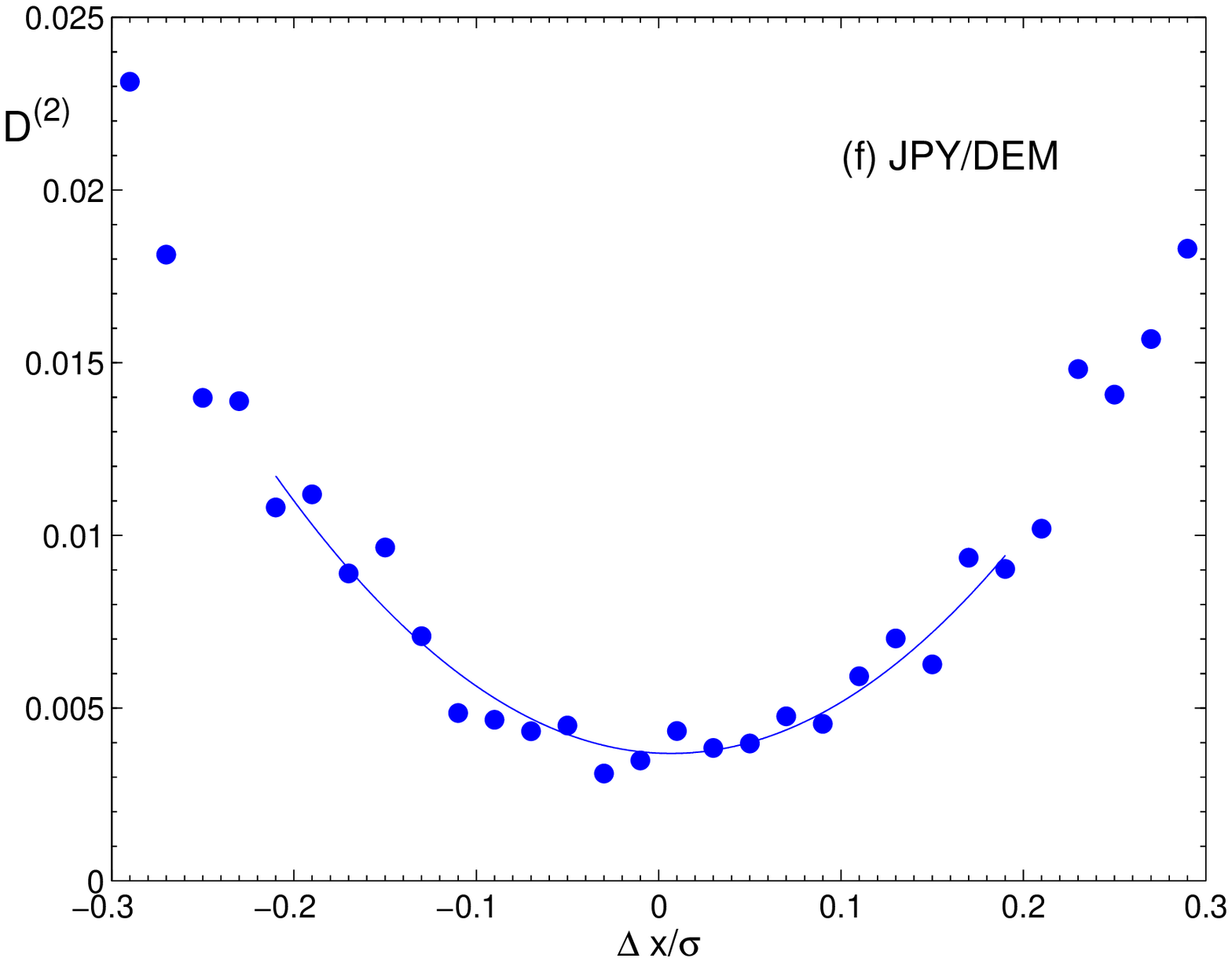} \vfill 
\includegraphics[width=.445\textwidth]{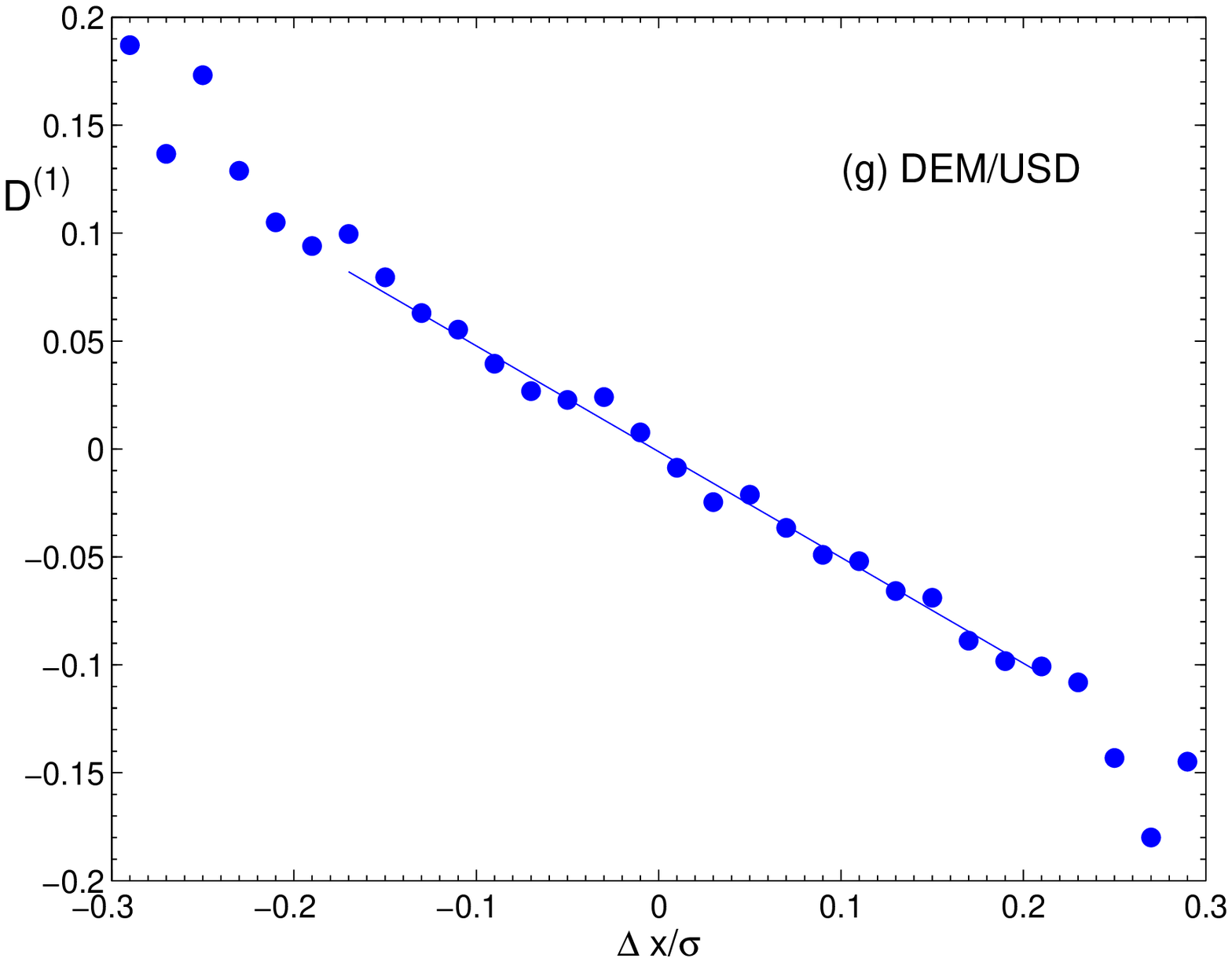} \hfill 
\includegraphics[width=.445\textwidth]{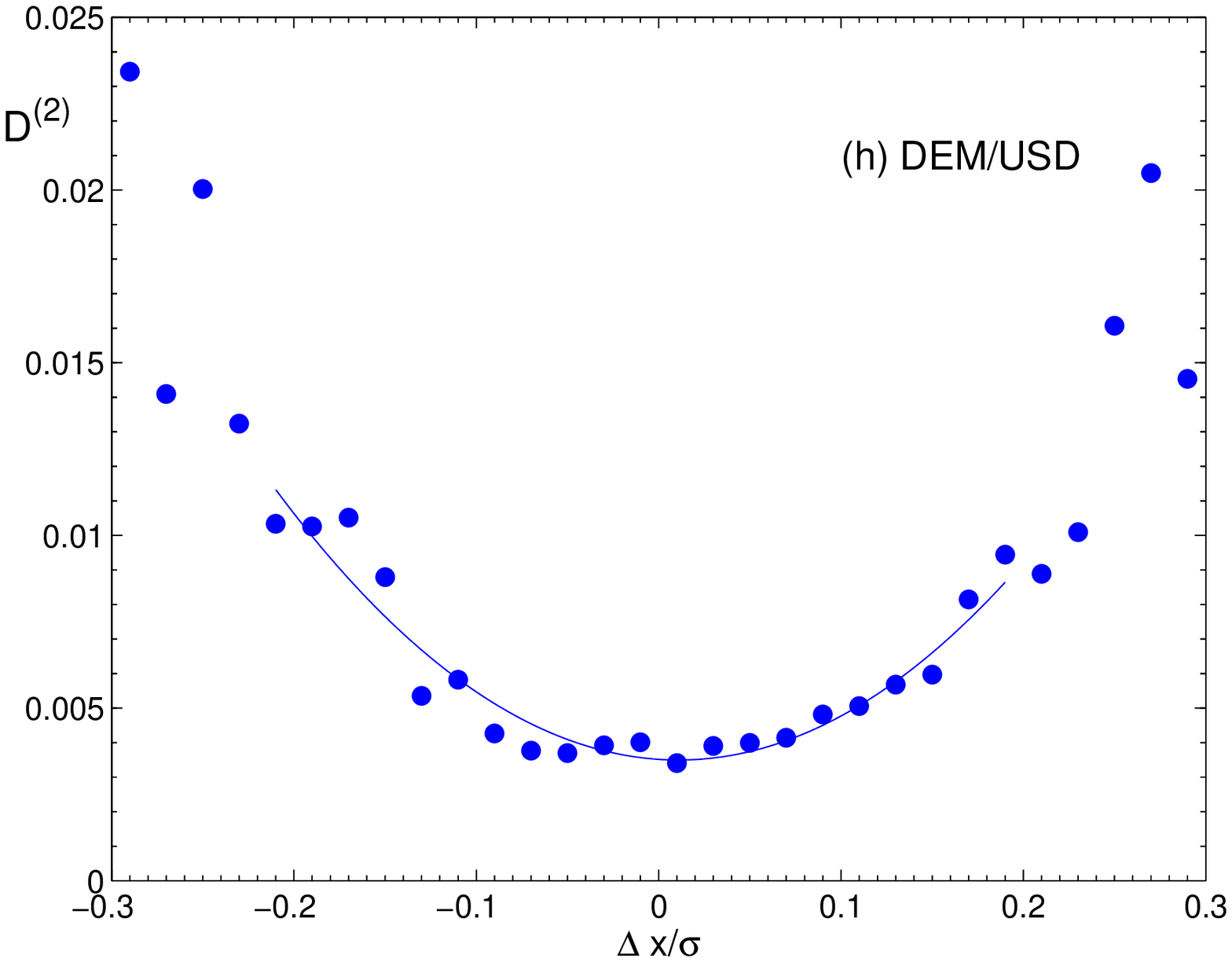} \caption{Functional 
dependence of the drift and diffusion coefficients $D^{(1)}$ and $D^{(2)}$ for 
the pdf evolution equation (3); $\Delta x$ is normalized  with respect to the 
value of the standard deviation $\sigma$ of the pdf increments at delay time 
32~days: (a,b) NIKKEI~225 and (c,d) NASDAQ closing price signal, (e,f) 
$JPY$/$DEM$ and (g,h) $DEM$/$USD$ exchange rates} \label{eps4} \end{figure}

\begin{figure} \centering \includegraphics[width=.48\textwidth]{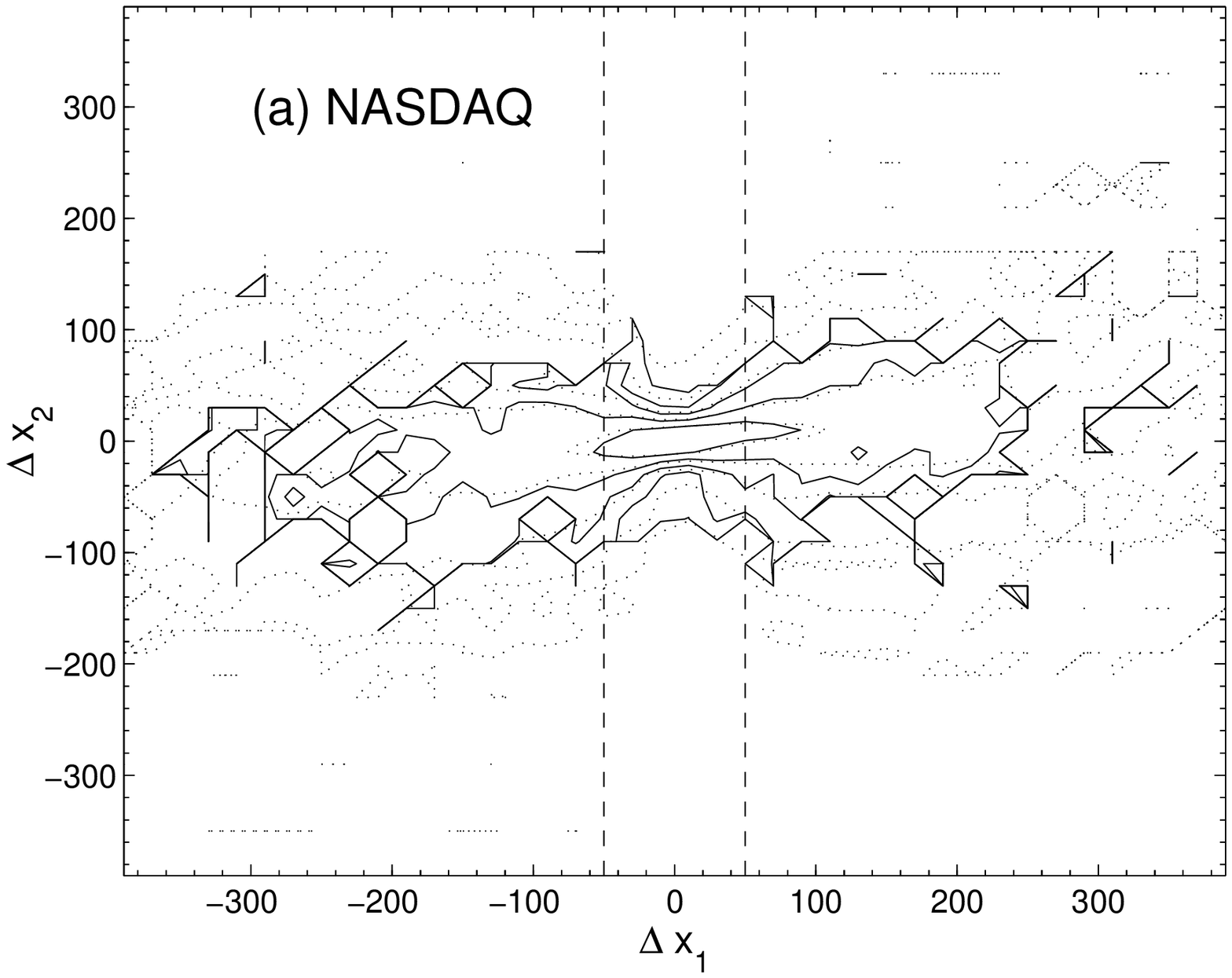} 
\vfill \includegraphics[width=.48\textwidth]{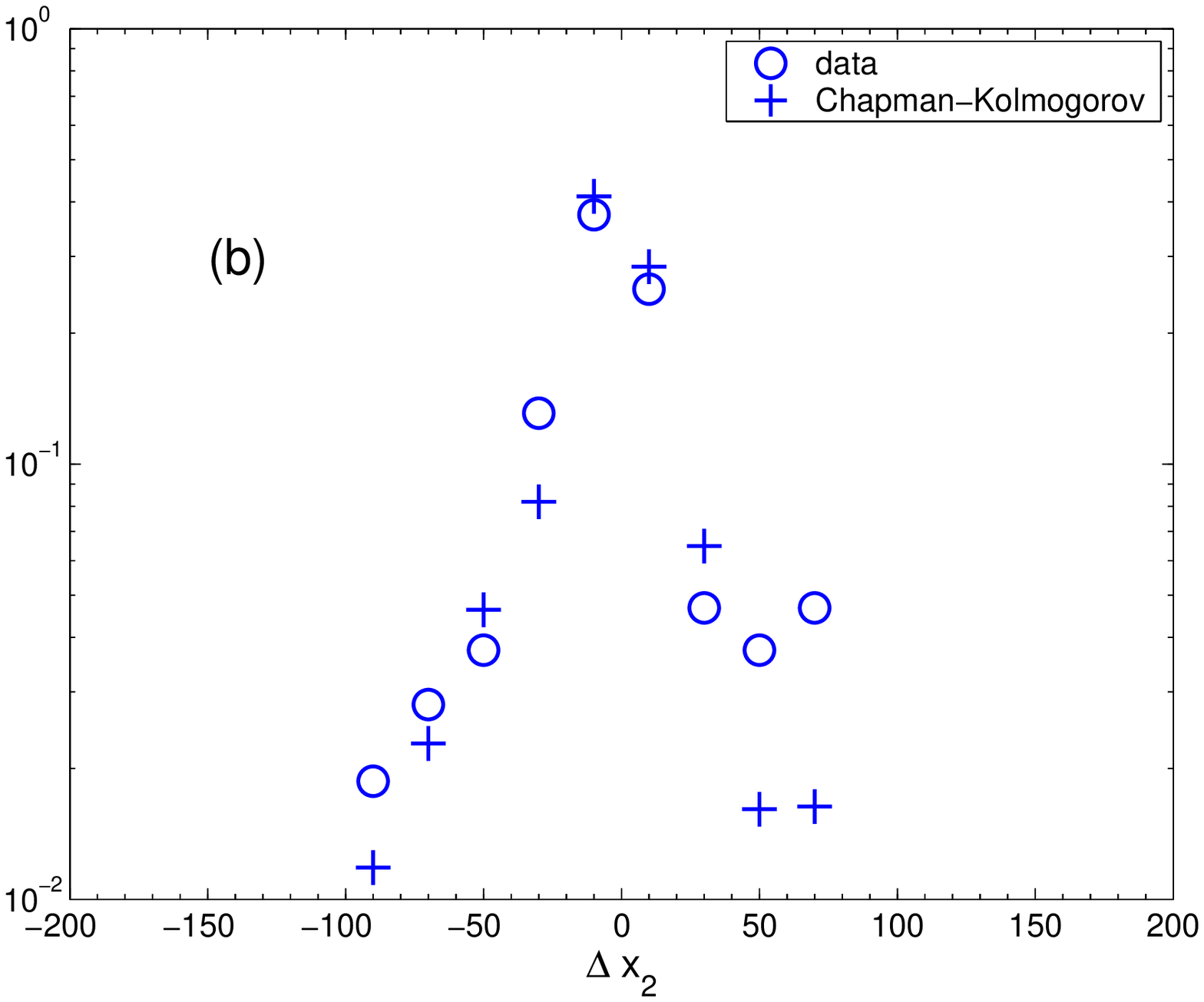} \hfill 
\includegraphics[width=.48\textwidth]{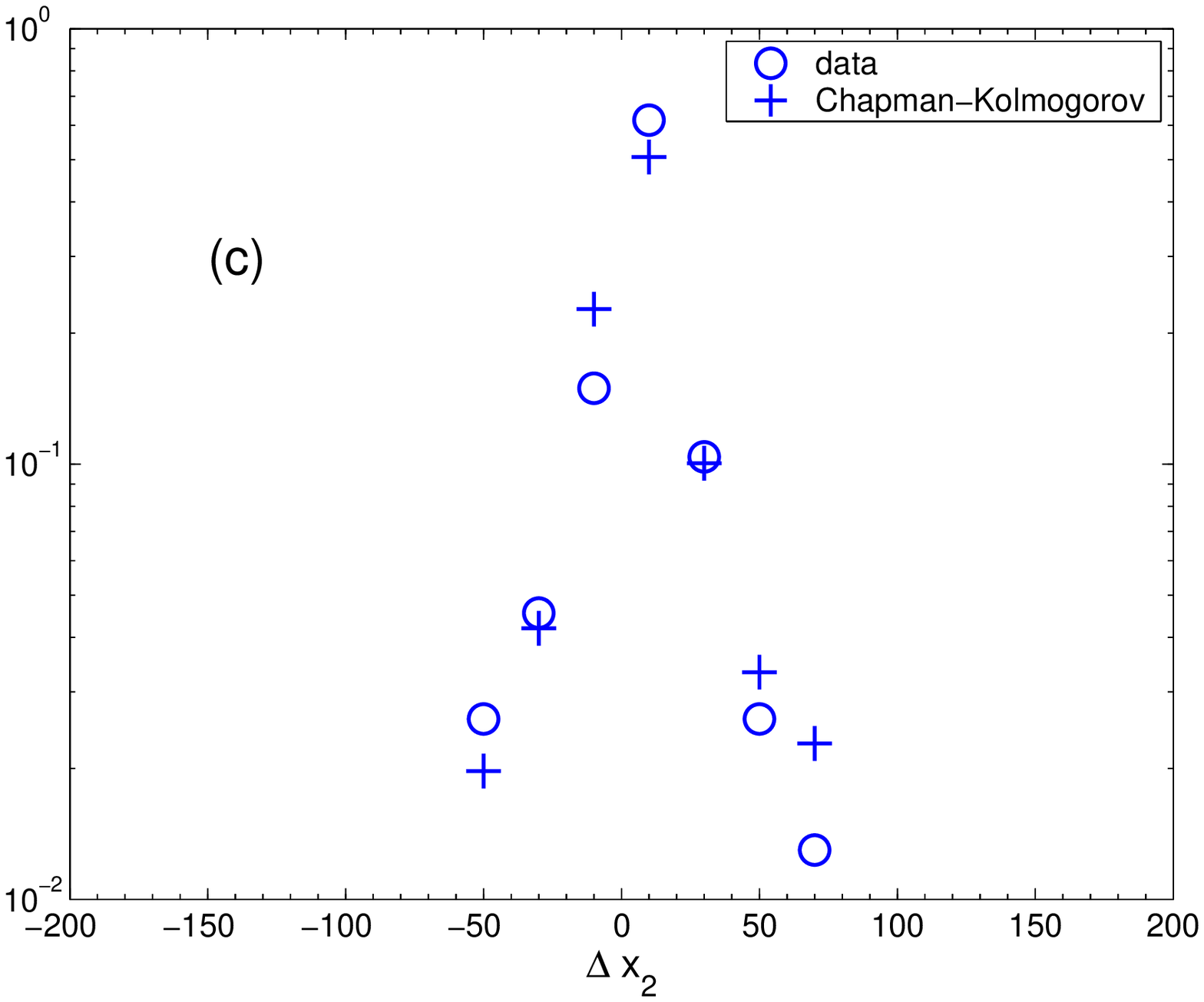} \caption{Equal probability 
contour plots of the conditional pdf $p(\Delta x_2,\Delta t_2|\Delta x_1,\Delta 
t_1)$ for two values of $\Delta t$, $\Delta t_1=8 \quad$ days, $\Delta t_2=1 
\quad$ day for NASDAQ. Contour levels correspond to 
$log_{10}p$=-0.5,-1.0,-1.5,-2.0,-2.5  from center to border; data (solid line) 
and solution of the Chapman Kolmogorov equation integration (dotted line); (b) 
and (c) data (circles) and solution of the Chapman Kolmogorov equation 
integration (plusses) for the corresponding pdf at $\Delta x_2$ = -50 and +50} 
\label{eps5} \end{figure}

\begin{table}[tbh] \centering \caption{Values of the polynomial coefficients 
defining the linear and quadratic dependence of the drift and diffusion 
coefficients $D^{(1)}= a_1(\Delta x/\sigma) + a_0$ and $D^{(2)}= b_2(\Delta 
x/\sigma)^2+ b_1(\Delta x/\sigma) + b_0$ for the FPE (3) of the normalized data 
series; $\sigma$ represents the normalization constant equal to the standard 
deviation of the $\Delta t$=32~days pdf} \renewcommand{\arraystretch}{1.4} 
\setlength\tabcolsep{5pt} \begin{tabular}{lccccccc} \hline 
%&\multicolumn{2-3}{$D^{(1)}$}&&\multicolumn{5-7}{$D^{(2)}$}&$\sigma$\\ 
&$a_1$&$a_0$&&$b_2$&$b_1$&$b_0$&$\sigma$\\ \hline NIKKEI~225& 
-0.54&0.002&&0.18&-0.004&0.003&1557.0 \\ NASDAQ&-0.49&-0.0004&&0.30&-0.010&0.001& 
198.11\\ $JPY$/$DEM$&-0.55&0.002&&0.17&-0.002&0.004&2.9111\\ 
$DEM$/$USD$&-0.49&-0.001&&0.16&-0.004&0.004&0.0808\\ \hline \end{tabular} 
\label{Tab1} \end{table}

The leading coefficient ($a_1$) of the linear $D^{(1)}$ dependence has 
approximately the same values for all studied signals, thus the same 
deterministic noise (drift coefficient). Note that the leading term ($b_2$) of 
the functional dependence of diffusion coefficient of the NASDAQ closing price 
signal is about twice the leading, i.e. second order coefficient, of the other 
three series of interest. This can be interpreted as if the stochastic component 
(diffusion coefficient) of the dynamics of NASDAQ is twice larger than the 
stochastic components of NIKKEI~225, $JPY$/$DEM$ and $DEM$/$USD$. A possible 
reason for such a behavior may be related to the transaction procedure on the 
NASDAQ. Our numerical result agrees with that of ref. \cite{friedrich} if a 
factor of ten is corrected in the latter ref. for $b_2$.

The validity of the Chapman-Kolmogorov equation has also been verified. A 
comparison of the directly evaluated conditional pdf with the numerical 
integration result (2) indicates that both pdf's are statistically identical. The 
more pronounced peak for the NASDAQ is recovered (see Fig. 5). An analytical form 
for the pdf's has been obtained by other authors \cite{Yako,Kozuki} but with 
models different from more classical ones \cite{Nekhi}.

\section{Conclusion}

The present study of the evolution of  Japan and US stock as well as foreign 
currency exchange markets has allowed us to point out the existence of 
deterministic and stochastic influences. Our results confirm those for high 
frequency (1 year long) data \cite{Bav,Kozuki}. The Markovian nature of the 
process governing the pdf evolution is confirmed for such long range data as in 
\cite{friedrich,Bav,Nekhi} for high frequency data. We found that the stochastic 
component (expressed through the diffusion coefficient) for NASDAQ is 
substantially larger (twice) than  for NIKKEI~225, $JPY$/$DEM$ and $DEM$/$USD$. 
This could be attributed to the electronic nature of executing transactions on 
NASDAQ, therefore to different stochastic forces for the market dynamics.

\vspace*{0.6cm}

{\noindent \large \bf Acknowledgements}

\vspace*{0.6cm}

KI and MA are very grateful to the organizers of the Symposium for their 
invitation and to the Symposium sponsors for financial support.

%INDEX%%%%%%%%%%%%%%%%%%%%%%%%%%%%%%%%%%%%%%%%%%%%%%%%%%%%%%%%%%%%%%% 
\clearpage \addcontentsline{toc}{section}{Index} \flushbottom \printindex 
\end{document}